%% file: main.tex
\documentclass[%
 reprint,
 amsmath,amssymb,
 aps,
floatfix,
]{revtex4-2}

\newcommand{\ii}{{\rm i}}
\newcommand{\ee}{{\rm e}}

\usepackage{graphicx}
\usepackage{dcolumn}
\usepackage{bm}

\usepackage{braket}
\usepackage[caption=false]{subfig}

\begin{document}


\title{Strong Correlations in the Dynamical Evolution of Lowest Landau Level Bosons }

\author{Y. Yang}
\author{N. R. Cooper}
\affiliation{T.C.M. Group, Cavendish Laboratory, University of Cambridge, J.J. Thomson Avenue, Cambridge CB3 0US, United Kingdom \looseness=-1}


\date{\today}

\begin{abstract}
Recent experiments with rotating Bose gases have demonstrated the interaction-driven hydrodynamic instability of an initial extended strip-like state in the lowest Landau level. We investigate this phenomenon in the low density limit, where the mean-field Gross--Pitaevskii theory becomes inadequate, using exact diagonalisation studies and analytic arguments. We show that the behaviour can be understood in terms of weakly-interacting repulsively-bound few-body clusters. Signatures of cluster behaviour are observed in the expectation values of observables which oscillate at frequencies characterised by the energies of few-body boundstates. Using a semiclassical theory for interacting clusters, we predict the long-time growth of the cloud width to be a power law in the logarithm of time. This slow thermalisation of bound clusters represents a form of quantum many-body scars.
\end{abstract}
\maketitle

\section{Introduction}
The behaviour of Bose--Einstein condensates (BECs) under rotation is drastically different from that of a normal fluid \cite{cooper_rapidly_2008, fetter_rotating_2009, viefers_quantum_2008, bloch_many-body_2008, matthews_vortices_1999, madison_vortex_2000, haljan_driving_2001, bradley_bose-einstein_2008, pethick_bose-einstein_2008, bretin_fast_2004, schweikhard_rapidly_2004, cooper_vortex_2005, yao_observation_2016}.
Rotating BECs host vortices carrying quantised vorticity, and at moderate rotation rates the ground state is a vortex lattice similar to those in type-II superconductors.
At very high rotation rates, theoretical studies predict that the vortex lattice is destabilised by quantum fluctuations, and the rotating Bose gas enters a strongly correlated regime \cite{wilkin_attractive_1998,cooper_quantum_2001}.
This regime is the bosonic analogue of the fractional quantum Hall effect (FQHE), featuring many phases and states familiar from the FQHE of electrons, including Laughlin states, composite fermions, and non-abelian phases.
Furthermore, recent experimental advances demonstrated the tantalising possibility of realising such states with ultracold atoms \cite{leonard_realization_2023}, paving the way for the quantum simulation of the FQHE.

Turning to the dynamical aspects of rotating BECs, a separate very interesting series of experiments were recently carried out \cite{yao_observation_2024, fletcher_geometric_2021, mukherjee_crystallization_2022}.
In these experiments, bosons were prepared in a rotating condensate of a strip-like shape.
This state was shown to be dynamically unstable and to spontaneously form a density modulation.
This behaviour was explained in both the lowest Landau level (LLL) regime and the hydrodynamic regime as a Kelvin--Helmholtz instability~\cite{mukherjee_crystallization_2022}.
On the other hand, given that these results were developed with either the Bogoliubov approximation or the mean-field Gross--Pitaevskii (GP) theory \cite{sinha_two-dimensional_2005}, one naturally wonders how the dynamics changes as the density is lowered, in the same way that the vortex lattice state transitions into the bosonic FQHE.

\begin{figure}[htb]
    \centering
    \includegraphics[width=0.85\linewidth]{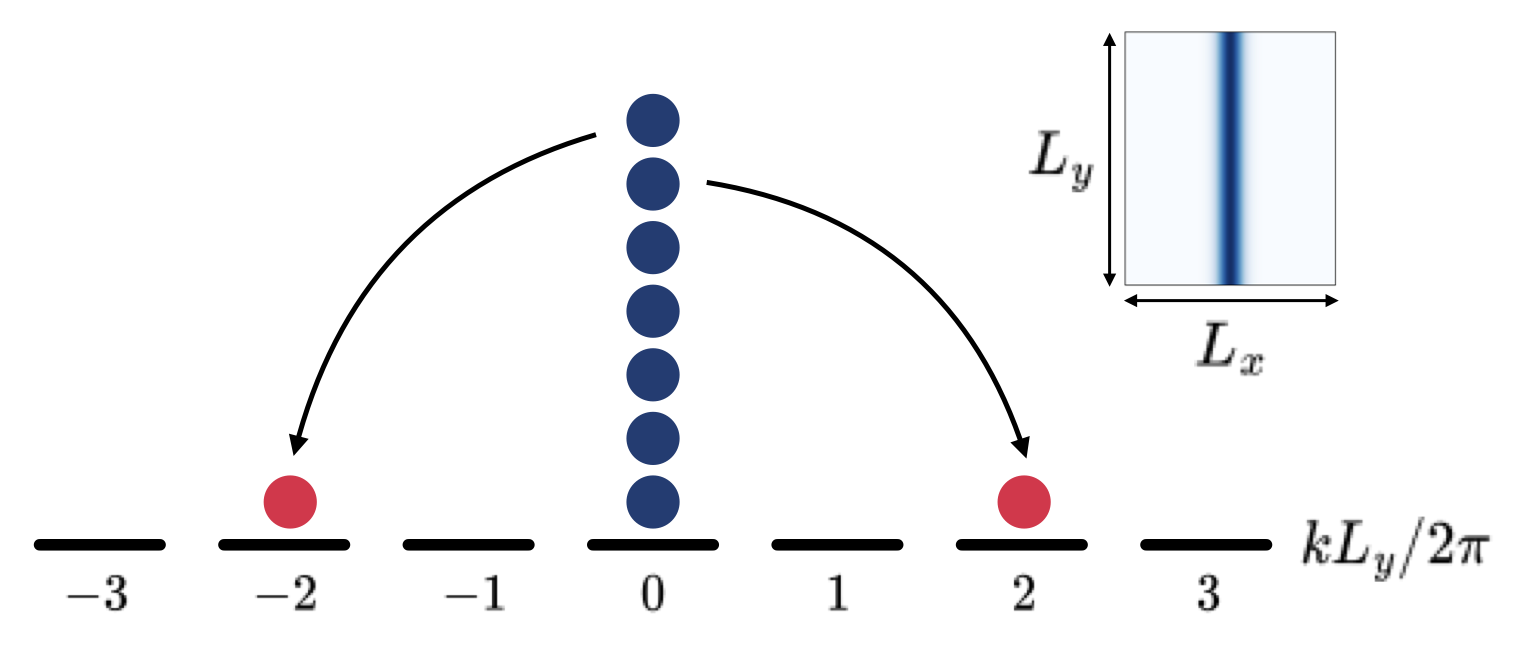}
    \caption{A schematic of the system in question. The initial state is the condensate (dark blue) in the zero momentum state in the Landau gauge in the LLL. Interactions cause the condensate to deplete, producing pairs of atoms in opposite momentum states (red). These give rise to the dynamic instabilities. The inset shows the initial state in real space on a torus of size $L_x\times L_y$.}
    \label{fig:demo}
\end{figure}

In this work, we investigate theoretically the dynamics of a strip-like condensate in this low density regime where GP theory becomes inadequate.
A schematic of the setup is shown in Fig.~\ref{fig:demo}, depicting the creation of pairs of particles with opposite momenta from the condensate due to the interactions between the bosons.
We use both exact diagonalisation (ED) methods and analytical approaches to obtain a consistent understanding of the key features.
We focus on observables that are most easily accessible in experiments: the particle density, the width of the cloud strip, and the density-density correlation function.
Interestingly, we find that these observables oscillate at frequencies that are inconsistent with the mean-field treatment.
Rather, as we explain, they can be understood in terms of a picture where the dynamics is dominated by few-body clusters: each cluster being a composite object of a few bosons with zero internal angular momentum.
We also show that the full excitation spectrum of contact interacting bosons in the LLL can be cast in a similar light, showing peaks in the densities of states at energies corresponding to non-interacting clusters.
Furthermore, we propose a semiclassical theory of cluster interactions, which explains further detailed structures of the density of states.
The semiclassical theory also predicts the width of the strip-like cloud to grow as a power law \textit{in the logarithm of time}.
This differs from diffusive or subdiffusive behaviours sometimes seen in the literature that are power laws \textit{in time itself} \cite{zerba_emergent_2025}.
We also show how the logarithmic timescales also apply to the thermalisation of the cloud.

\section{System and Observables}
We start with the second quantised Hamiltonian for dilute Bose gases, with field operators $\hat{\psi}^{(\dag)}(\mathbf{r})$. We consider particles of mass $M$ in two dimensions and impose an effective magnetic field via minimal substitution $\mathbf{p}\to\mathbf{p}+e\mathbf{A}$~\cite{cooper_rapidly_2008}.
We choose the Landau gauge where $\mathbf{A}=(0,Bx)$ so that the Hamiltonian becomes
\begin{multline}
    \hat{H} = \int\mathrm{d}^2r\,\biggl\{\hat{\psi}^\dagger(\mathbf{r})\biggl[\frac{p_x^2}{2M}+\frac12 M\omega_c^2(x+k_y\ell_B^2)^2\biggr]\hat{\psi}(\mathbf{r})+ {}\\
    \frac{g}{2}\hat{\psi}^\dagger(\mathbf{r})\hat{\psi}^\dagger(\mathbf{r})\hat{\psi}(\mathbf{r})\hat{\psi}(\mathbf{r})\biggr\}
\end{multline}
where we define the cyclotron frequency $\omega_c\equiv eB/M$ and the magnetic length $\ell_B\equiv\sqrt{\hbar/eB}$. (This describes harmonically trapped gases rotating at the trap frequency $\Omega$ when $\omega_c = 2\Omega$~\cite{cooper_rapidly_2008}.)
$g$ is the strength of contact interactions between the bosons.
For the rest of this paper, we work in units $\hbar=1$.

The single-particle part of the Hamiltonian is that of a harmonic oscillator, giving quantised Landau levels $E_n=\omega_c(n+1/2)$ and eigenstates labelled by momentum $k$ along the $y$-direction
    $\phi_{nk}(\mathbf{r})\propto \ee^{-(x/\ell_B+k\ell_B)^2/2} H_n(x/\ell_B+k\ell_B)\ee^{\ii k y}$,
where $H_n$ are Hermite polynomials.
Imposing periodic boundary conditions (PBCs) along the $y$-direction gives quantised values of $k=2\pi m/L_y$ for integer $m$.
PBCs along the $x$-direction leads to a finite number of states $N_\phi=L_xL_y/2\pi\ell_B^2$ per Landau level, which is also the number of magnetic flux quanta threaded through the gas~\cite{yoshioka_ground_1983}.

Assuming that the cyclotron energy is larger than the interaction energy (i.e., very fast rotation), the gas will reside fully in the LLL ($n=0$) \cite{wilkin_attractive_1998,cooper_rapidly_2008,lewin_strongly_2009}.
Projecting the field operators onto the LLL and accounting for the periodicity, \textit{i.e.} $\hat{\psi}(\mathbf{r})\to\sum_{k} \hat{a}_k \bigl[ \sum_m \phi_{0,k+mN_\phi}(\mathbf{r}) \bigr]$ \cite{yoshioka_ground_1983}, we get a Hamiltonian
\begin{equation}\label{eq:ham}
    \hat H=\frac{\ell_B}{L_y}\sum_{k_1k_2k_3k_4} V_{k_1k_2k_3k_4} \hat a_{k_1}^\dagger \hat a_{k_2}^\dagger \hat a_{k_3} \hat a_{k_4}.
\end{equation}
where the potential energy is given by 
\begin{multline}
    V_{k_1k_2k_3k_4}=\frac{g}{2\sqrt{2\pi}\ell_B^2} \sum_{m_1m_2m_3m_4}
    \delta_{q_1+q_2-q_3-q_4} \times {}\\
    \exp\Biggl[-\frac{\ell_B^2}2\sum_{i=1}^4 q_i^2+\frac{\ell_B^2}8\biggl(\sum_{i=1}^4 q_i \biggr)^2\Biggr].
\end{multline}
The $\delta$ is the Kronecker delta and $q_i$ is a shorthand for $q_i\equiv k_i+m_i N_\phi$.
The kinetic energy in the LLL is a constant and has been removed from the Hamiltonian.
These matrix elements provide the scattering on the torus arising from the contact interactions (as seen in Fig.~\ref{fig:demo}), the strength of which is conveniently represented by the Haldane pseudopotential for two particles with zero relative angular momentum, $V_0=g/4\pi\ell_B^2$.
The delta function conserves the total momentum modulo the additions of integer multiples of $2\pi N_\phi/L_y = L_x/\ell_B^2$ \cite{haldane_many-particle_1985}.

Our main numerical approach is the exact diagonalisation of this Hamiltonian for $N$ particles and $N_\phi$ flux quanta.
We also study the time evolution of the strip-like condensate $\ket{\Psi(t=0)}\propto (\hat a_0^\dagger)^N\ket{0}$ by integrating Schr\"odinger's equation $\ii \partial_t\ket{\Psi(t)}=\hat{H}\ket{\Psi(t)}$.

To analyse the numerical results, we choose the occupation numbers for each momentum state $\hat\rho_k\equiv \hat a_k^\dagger \hat a_k$ and their correlations as our observables.
Additionally, we study the squared cloud width $\hat x^2/\ell_B^2\equiv \bigl(\frac12+\frac1N\sum_k (k\ell_B)^2\hat \rho_k\bigr)$, which helps to clarify the long-term behaviour of the cloud strip.

\section{Overview of Numerical Results}

We first discuss the energy scales and densities determining the validity of mean-field descriptions: one expects GP theory to break down when the density of the condensate is sufficiently low and quantum effects dominate.

In equilibrium systems, the 2D density $n_\text{2D}$, and equivalently the filling fraction $\nu=2\pi n_\text{2D}\ell_B^2$, are used as control parameters of the system.
For instance, in studies of the ground state, it is known that the uncertainty in the position of a vortex is $\propto 1/\sqrt{n_\text{2D}}$.
In a low density regime where these quantum fluctuations become comparable to the vortex spacing, or $\nu \sim n_\text{2D}\ell_B^2 \lesssim \mathcal{O}(1)$, the vortices can no longer be treated as classical point-like objects as in GP theory \cite{cooper_rapidly_2008}, instead transitioning into a strongly correlated regime \cite{cooper_quantum_2001}.

In out-of-equilibrium systems, the density and the filling fraction are neither homogeneous nor constant, so they cannot be the sole control parameters of the system.
Instead, we must also consider timescales or energy scales.
In the mean field GP theory, the parameter used is $gn_\text{2D}$, which is the typical energy per particle when the density is $n_\text{2D}$.
However, when the gas becomes so dilute that each particle interacts only with a few others (again $n_\text{2D}\ell_B^2 \lesssim \mathcal{O}(1)$), the granularity becomes important.
Indeed, the zeroth Haldane pseudopotential $V_0$ is the fundamental energy scale of two-body interactions.
Thus, in the low density regime where the mean field energy $gn_\text{2D}$ becomes comparable to $V_0$, or equivalently when $n_\text{2D}\ell_B^2 \lesssim \mathcal{O}(1)$, GP results can no longer be expected to hold.
Instead, the exact dynamics is governed by both $V_0$ and the density $n_\text{2D}$ \textit{separately}.
Note that, when discussing the evolution of the strip-like condensate, the density is more conveniently expressed in terms of a 1D filling fraction $\nu_\text{1D}\equiv N\ell_B/L_y$ along the strip, which is also the initial density of the cloud.

Interestingly, these two considerations lead to essentially the same parameter: GP will not hold either in- or out of equilibrium when the density is so dilute that the typical particle spacing locally cannot be considered to be much less than the magnetic length.

\begin{figure}[ht]
    \centering
    \subfloat[]{\includegraphics[width=0.75\linewidth]{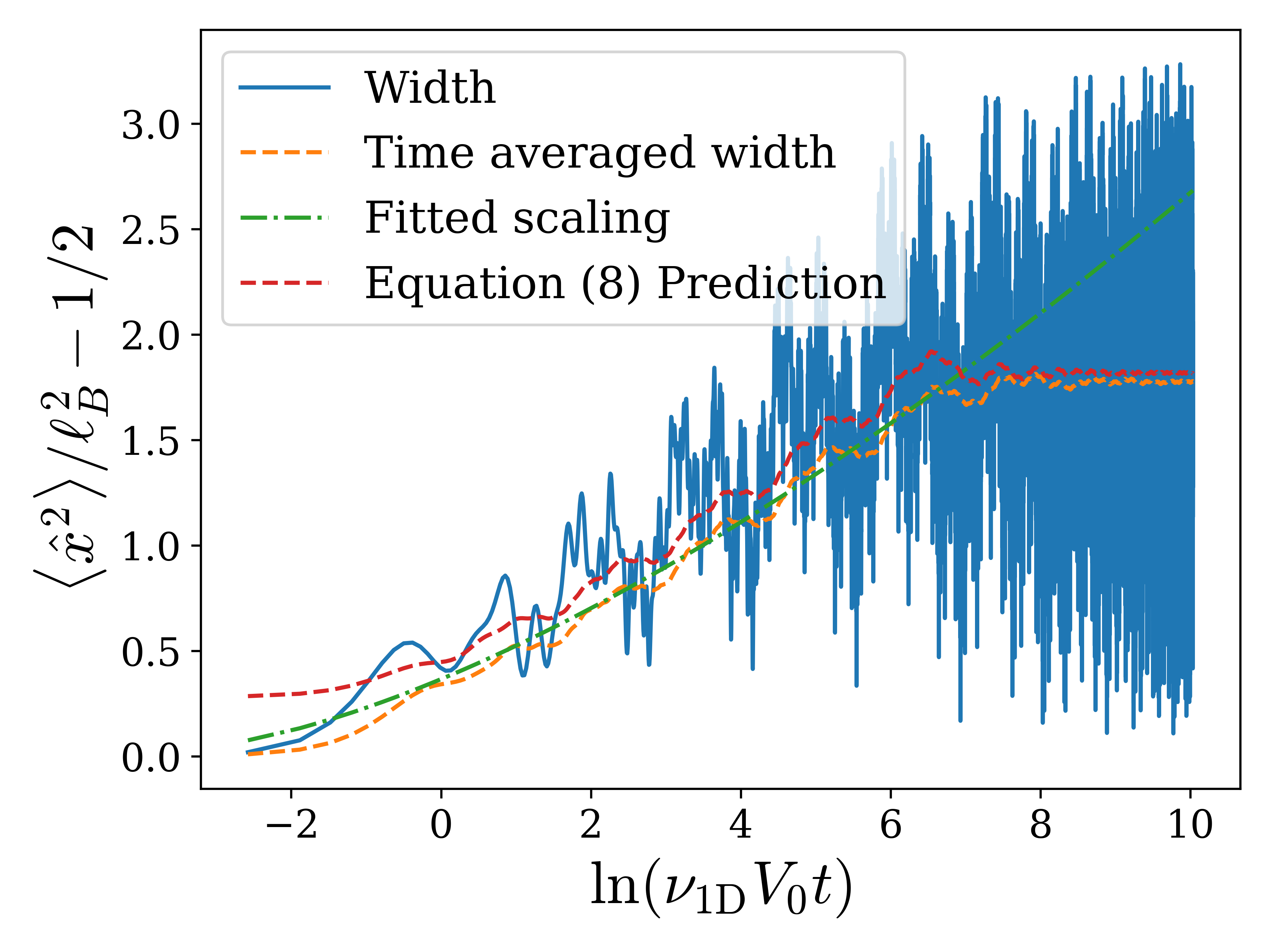}}\\
    \subfloat[]{\includegraphics[width=0.75\linewidth]{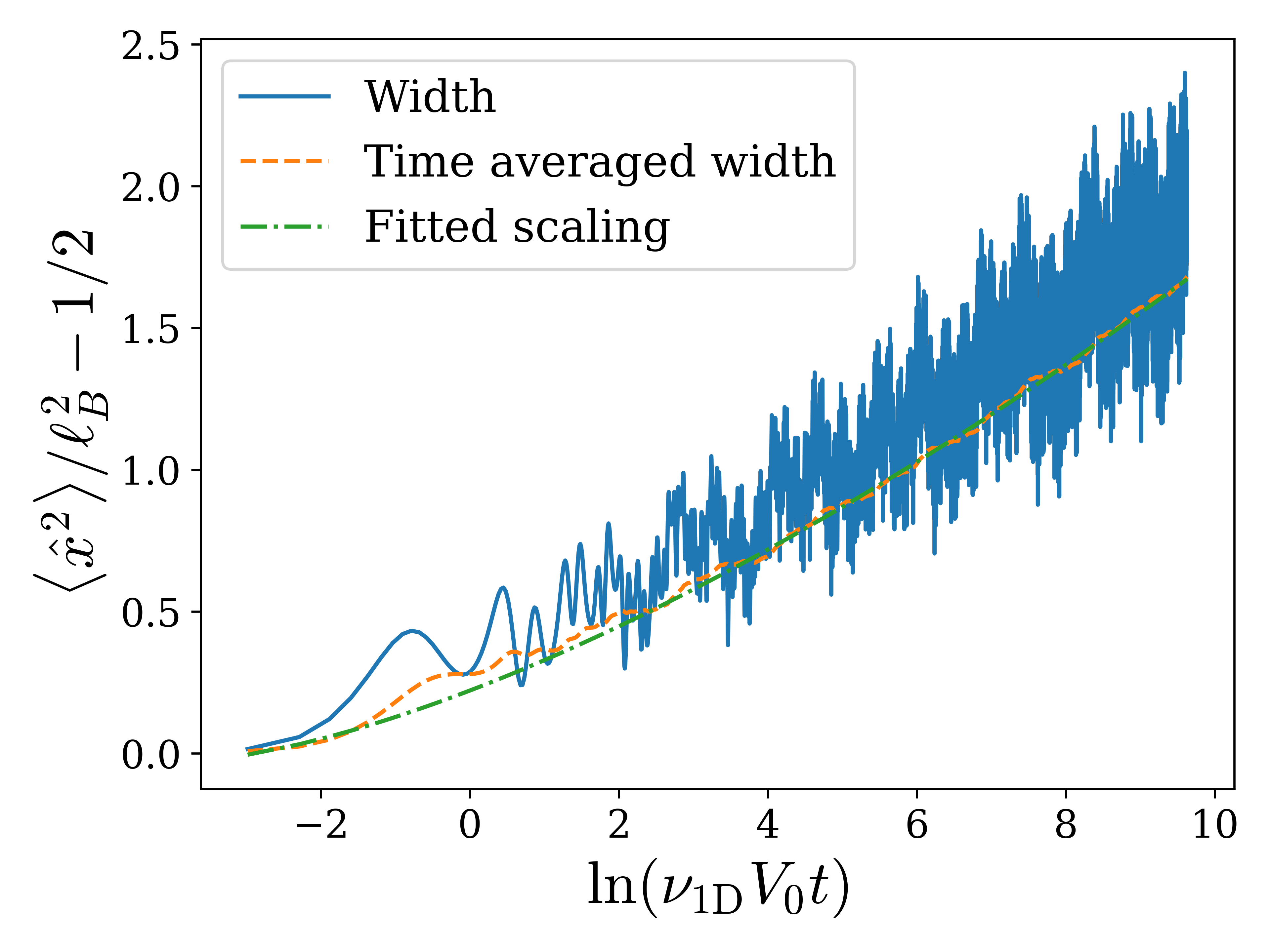}}
    \caption{The squared cloud width in (a) $N=3, \nu_\text{1D}=0.24$ and (b) $N=4, \nu_\text{1D}=0.16$ systems. The blue is the raw squared width, showing fast oscillations. The dashed orange curve is the time-averaged width, showing a long term growth that is consistent with our $(\ln t)^{1.5}$ prediction (dash-dotted green curve) described in section~\ref{sec:growth}. The scaling is obtained from numerical fits. In $N=3$ we have direct calculations based on equation~\ref{eq:sinc} yielding the dashed red curves. }
    \label{fig:width_sinc}
\end{figure}

We present the simulated squared cloud width in Fig.~\ref{fig:width_sinc} for two systems, which are representative of the low-density (strongly correlated) regime for which GP theory is inadequate.
For the system to qualify as low-density, the energy per particle $g(N-1)/2\sqrt{2\pi}L_y\ell_B$ must be comparable to $V_0$, so $\nu_\text{1D} \lesssim 1/\sqrt{2\pi}\approx 0.40$.
In the two examples, $\nu_\text{1D}=0.24$ and $0.16$, respectively.
Note that a logarithmic scale is used for time, while the width is displayed on a linear scale.

There are several interesting features to analyse and understand.
\begin{itemize}
    \item The cloud width ($\braket{\Psi(t)| \hat{x}^2 | \Psi(t)}$, blue curve) shows rapid oscillations in time.
\item 
These rapid oscillations can be removed with time-averaging (dashed orange curve), where $\overline{\langle \hat x^2\rangle}(T)\equiv \frac{1}{T}\int_0^T \braket{\Psi(t)| \hat{x}^2 | \Psi(t)}\,\mathrm{d}t$.
The time-averaged width shows periodic undulations in the logarithm of time while also showing an overall growth. This is particularly clear for the $N=3$ system (with a period of $\approx 1.4$), Fig.~\ref{fig:width_sinc}(a), but also present in all cases.

\item There is an eventual saturation of the cloud width due to finite size effects in the system. This is evident for $N=3$, but not in $N=4$ for the timescale shown in Fig.~\ref{fig:width_sinc}(b).

\end{itemize}

Here, we propose a theory of repulsively-bound few-boson clusters that explains all of these features.

\section{Non-interacting Clusters}\label{sec:non_interact}

A set of $c$ contact-interacting bosons in the LLL always have an exact eigenstate with zero angular momentum (all $c$ particles overlap completely) and energy $V_0 c(c-1)/2$. 
This is what we refer to as a cluster of size $c$: an object composed of a few particles and zero internal angular momentum.

Exact diagonalisation studies have also shown that excited states can be viewed as being composed of separate clusters of this type.
An example is the $N=3$ three-body system, in which exact eigenstates feature a pair (a cluster of size $c=2$) and a singlet (a cluster of size $c=1$) and their energies approach $V_0$ as their angular momentum increases~\cite{mashkevich_exact_2007}.



In the following sections, we will begin with a model of the $N$-particle systems as a set of non-interacting clusters.
Concretely, a state with a definite cluster composition---a cluster state---satisfies
\begin{equation}\label{eq:decomp}
    N=\sum_{c}cn_c,\qquad E=\sum_c n_c\frac{c(c-1)}{2}V_0,
\end{equation}
where $E$ is the total energy and $n_c$ is the number of $c$-clusters.
Each set of cluster numbers $\{n_c\}$---with $c=1,2,3,\ldots $ representing what we will refer to as ``singlets'', ``doublets'' (also ``pairs''), ``triplets'', etc.---describes a distinct decomposition, and we denote the cluster state by $\ket{\{n_c\}}$.

\subsubsection{Fast Oscillations}\label{sec:obs}

Within the cluster picture, the fast oscillations of the cloud width arise from the fact that the initial state is a superposition of cluster states of different $\{n_c\}$, and therefore of different energies (equation~\ref{eq:decomp}).

\begin{figure}[htb]
    \centering
    \includegraphics[width=0.75\linewidth]{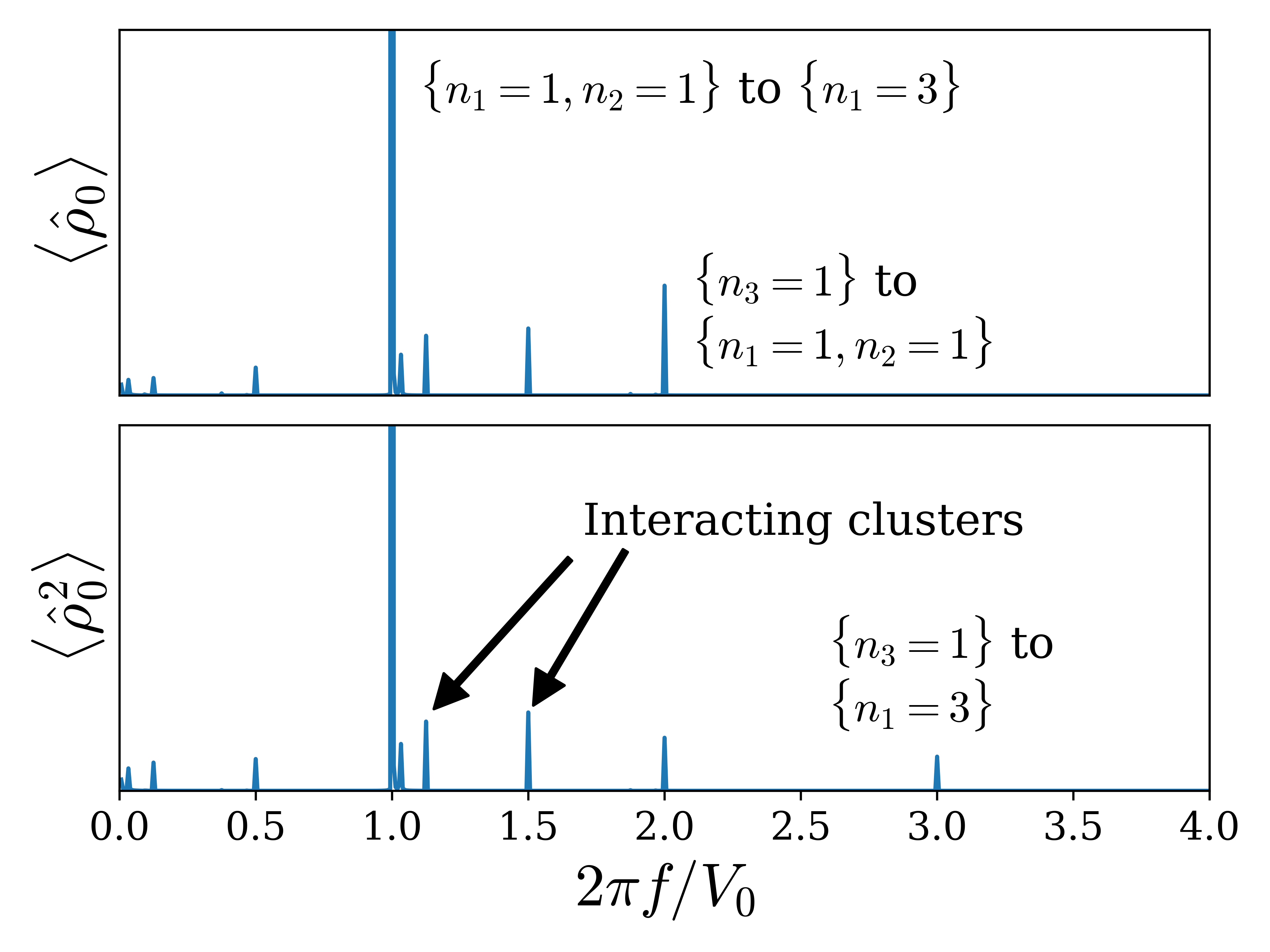}
    \caption{The power spectra of $\langle\hat\rho_0 \rangle$ and $\langle\hat\rho_0^2 \rangle$ in the $N=3, \nu_\text{1D}=0.12$ system. Peaks corresponding to transitions between different cluster states can be seen, including the $3V_0\to V_0$ transition. The $3V_0\to 0$ transition is missing in $\langle\hat\rho_0 \rangle$ but is present in the $\langle\hat\rho_0^2 \rangle$ spectrum.}
    \label{fig:vis}
\end{figure}

The power spectra of the central density $\langle\hat\rho_0\rangle$ and the density squared $\langle\hat\rho_0\hat\rho_0\rangle$ are shown in Fig.~\ref{fig:vis}, showing sharp peaks in frequency space.
The prominent frequency peak corresponds to the energy of the repulsively bound two-particle cluster, which is given by the Haldane pseudopotential $2\pi f = V_0$.
Since this characterises the energy of a completely overlapping pair of particles, it suggests that the relevant state has one pair of bosons and one isolated singlet, \textit{i.e.}, $\ket{\{n_1=1, n_2=1\}}$.
Some other peaks can be explained as differences between cluster energies.
For instance, the peak at $2V_0$ is obtained as the difference between the aforementioned $\{n_1=1, n_2=1\}$ state and the $\{n_3=1\}$ state, whose energy is $c(c-1)V_0/2=3V_0$ for $c=3$.
These observations give strong evidence that the dynamics of the cloud strip should be described in terms of weakly-interacting few-body clusters.
Clusters account for the main peaks at integer multiples of $V_0$, while others must be explained with cluster-cluster interactions that we explore in Section~\ref{sec:interact}.

Curiously, despite the above interpretation of the $2V_0$ peak implying the existence of the $\{n_3=1\}$ state, the $3V_0$ peak itself is not observed.
This is because of ``selection rules'' that eliminate certain frequencies from the spectra.
Consider the $V_0$ peak that comes from the matrix element $\braket{\{n_1=3\}|\hat\rho_k|\{n_1=1, n_2=1\}}$.
The density operator has one creation and one annihilation operator each, so it can ``break apart'' the pair.
The matrix element is thus non-zero, giving the transition frequency of $V_0$.
In contrast, the matrix element governing the $3V_0$ transition $\braket{\{n_1=3\}|\hat\rho_k|\{n_3=1\}}=0$ because two particles must be moved to break up the triplet, but only one annihilation operator is available.
Hence, we do not see a frequency peak at $3V_0$.
On the other hand, when we examine the square of the densities, the $3V_0$ peak is present, because $\hat\rho_k^2$ can move two particles.
In other words, the matrix element $\braket{\{n_1=3\}|\hat\rho_k^2|\{n_3=1\}}\neq 0$.

The $N=4$ spectra tell a similar story, with a $3V_0$ peak present in the $\langle\hat\rho_0 \rangle$ spectrum due to the $\{n_4=1\}$ to $\{n_1=1,n_3=1\}$ transition.
The $4V_0$ and $5V_0$ peaks are present in the $\langle\hat\rho_0^2 \rangle$ spectrum due to the $\{n_4=1\}$ to $\{n_1=2,n_2=1\}$ and $\{n_2=2\}$ transitions.
The $6V_0$ peak is absent in both spectra.

These selection rules further support the cluster picture, as shown in the lower panel of Fig.~\ref{fig:vis}.

\subsubsection{Energy and energy density correlator spectra}
Given that the dynamical features of the cloud's time evolution are explained from a cluster perspective, it is natural to ask if all eigenstates of the Hamiltonian are cluster states.

\begin{figure}[ht]
    \centering
    \subfloat[\label{sfig:dos}Density of states]{
        \includegraphics[width=0.75\linewidth]{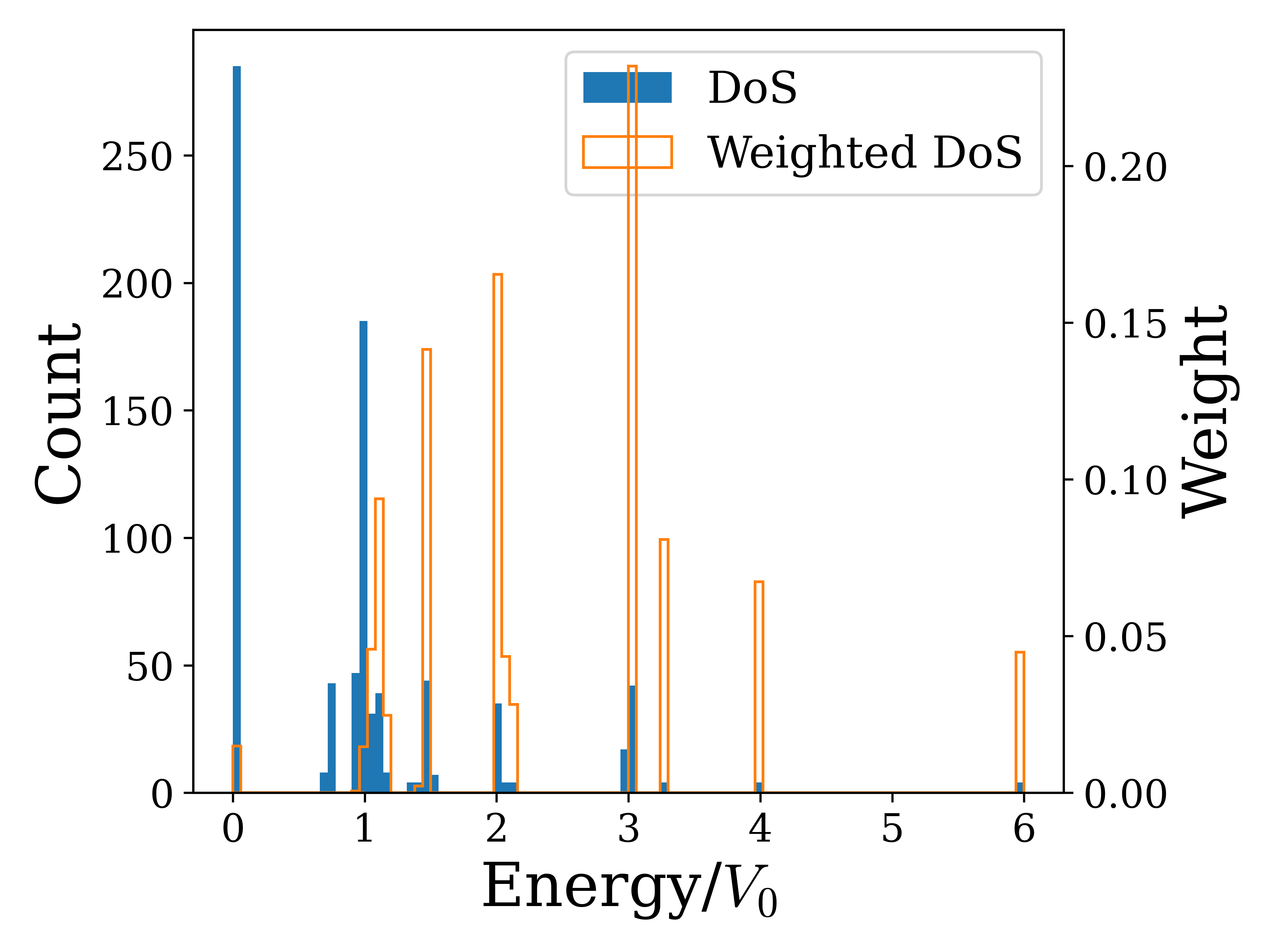}    
    }\\
    \subfloat[\label{sfig:eigval}Comparison of eigenvalues]{
        \includegraphics[width=0.75\linewidth]{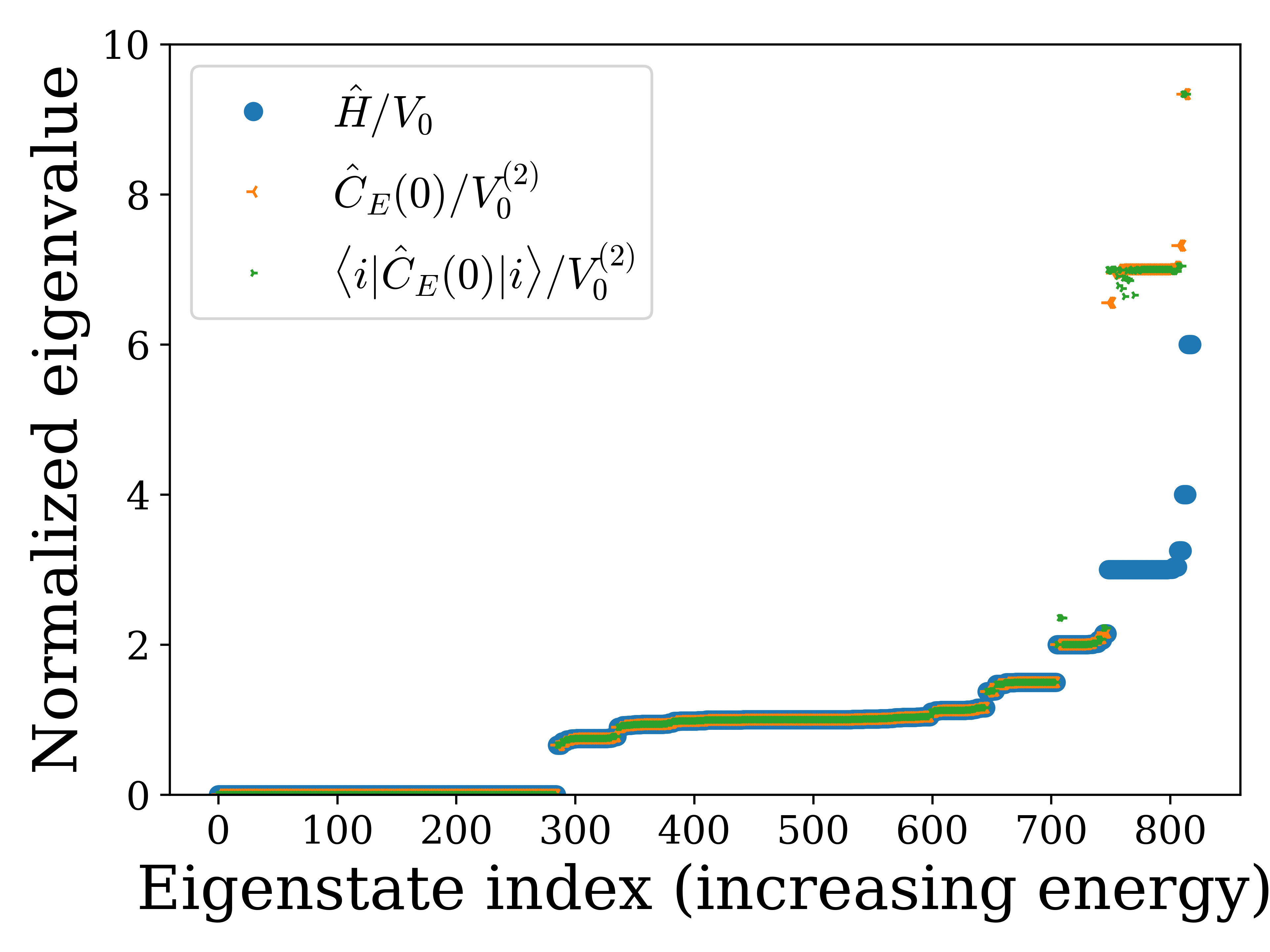}
    }
    \caption{The density of states and the normalized eigenvalues of the Hamiltonian and $\hat C_E(0)$ in $N=4, \nu_\text{1D}=0.32$. In (a), the solid blue shows the density of states, while the hollow orange shows the density of states weighted by the their overlaps with the initial state $|\braket{n|\Psi(t=0)}|^2$. In (b), the $\hat C_E(0)$ eigenvalues correspond well to the energy eigenvalues in the manner of equation~\ref{eq:decomp}: energy eigenstates $\ket{i}$ are nearly eigenstates of $\hat C_E(0)$ as $\braket{i|\hat C_E(0)|i}$ closely reproduces the $\hat C_E(0)$ eigenvalues. In particular, the $2V_0$ states are precisely the $2V_0^{(2)}$ states. The $25V_0^{(2)}$ states are present but not shown.}
    \label{fig:dos}
\end{figure}

In the density of states obtained via ED, the major peaks can be identified with the decompositions in equation~\ref{eq:decomp}, \textit{e.g.}, in Fig.~\ref{sfig:dos}.
There are peaks at energies $V_0,3V_0,6V_0$, corresponding to pairs ($\{n_1=2, n_2=1\}$), triplets ($\{n_1=1, n_3=1\}$), and quadruplets ($\{n_4=1\}$) of particles.
The $2V_0$ peaks can be interpreted as two pairs ($\{n_2=2\}$), a possibility in a $N=4$ system.

Furthermore, given equation~\ref{eq:decomp}, the expectation value of local observable $\hat{O}$ in a state should be $\sum_c n_c \braket{c|\hat{O}|c}$ where $\ket{c}\propto (\hat\psi^\dagger)^c\ket{0}$.
As a demonstration, we calculate the values of the (zero-separation) energy density correlations $\hat C_E(0)\equiv \int\mathrm{d}^2r\, \bigl( \frac{g}{2} \hat\psi^\dagger\hat\psi^\dagger\hat\psi\hat\psi\bigr)^2$.
Denoting the expectation value in a pair state as $V_0^{(2)}\equiv V_0^2/\pi\ell_B^2$, we find by explicit computation the expectation values in triplets to be $7V_0^{(2)}$ and those in quadruplets to be $25V_0^{(2)}$.
Importantly, in a state of two pairs, the expectation value $\langle \hat C_E(0) \rangle$ should be $2V_0^{(2)}$, since we have $n_2=2$.

These predictions are confirmed numerically.
In Fig.~\ref{sfig:eigval}, the states with energy  around $2V_0$ (shown in blue, index from 700 to 740), which we identify to be $\{n_2=2\}$ states, have $\langle \hat C_E(0)\rangle= 2V_0^{(2)}$ (shown in green).
Similarly, the $3V_0$ states (index from 740 to 800), which we identify to be $\{n_1=1, n_3=1\}$, have $\langle \hat C_E(0)\rangle= 7V_0^{(2)}$, as expected for the triplet.

This shows that many energy eigenstates can indeed be considered as combinations of independent clusters.

\section{Interacting Clusters}\label{sec:interact}
In the preceding sections, we showed how the energy spectra of a rotating Bose gas in the LLL with contact interactions can be deciphered with a picture of few-body clusters.
This explains the fast oscillations in Fig.~\ref{fig:width_sinc}.
Nonetheless, there are eigenstates whose energies cannot be explained in terms of energy differences of the non-interacting clusters (equation~\ref{eq:decomp}).
To account for these, we must now include interactions between clusters.
In fact, non-interacting clusters have no dynamics beyond the rapid oscillatory behaviour described previously.
In particular, there could be no gradual expansion of the cloud: if all energies are multiples of $V_0$, the system must be periodic in time with a period $2\pi/V_0$.
In this section, we construct a simple theory for the interactions between the clusters and use it to predict the long-term dynamics of the cloud.

\subsubsection{Classical and semiclassical cluster interactions}
We start by constructing a classical Hamiltonian from the interaction energy between clusters that captures their dynamics at low densities.
The interaction energy between two singlets positioned at $\mathbf{r}$ and $\mathbf{r}'$ can be shown to be $V_0\operatorname{sech}[(\mathbf{r}-\mathbf{r}')^2/4\ell_B^2]$, while that between two larger clusters can be approximated by multiplying this formula by the sizes of the clusters.
Importantly, the interaction energy is Gaussian $\sim V_0\ee^{-r^2/4\ell_B^2}$ as $|\mathbf{r}-\mathbf{r}'|\to\infty$.

We focus on interactions between pairs of clusters in subsequent analyses, since they are dominant in the low-density limit we are considering and are sufficient to explain the numerical results presented in Fig.~\ref{fig:width_sinc}.
While interactions between three or more clusters are in principle present, they merely produce higher order corrections and add fine structure to pairwise interactions.

From the interaction energy, we see classically that the dynamics depends on the logarithm of time.
Consider two clusters separated by a distance $r$.
The force between them is $F=-\nabla E\sim V_0r\ee^{-r^2/4\ell_B^2}/\ell_B^2$ for large $r$, and the velocity due to the force is $F/eB\sim V_0r\ee^{-r^2/4\ell_B^2}/\hbar$.
For the motion to be appreciable on the scale of $r$, a time of $V_0t/\hbar \sim \ee^{r^2/4\ell_B^2}$ is required, meaning $r^2/\ell_B^2\sim \ln(V_0t/\hbar)$. Thus a dependence on the {\it logarithm} of time arises naturally in the classical analysis.
We shall see later how this emerges in the quantum picture, too.

\begin{figure}[ht]
    \centering
    \includegraphics[width=0.75\linewidth]{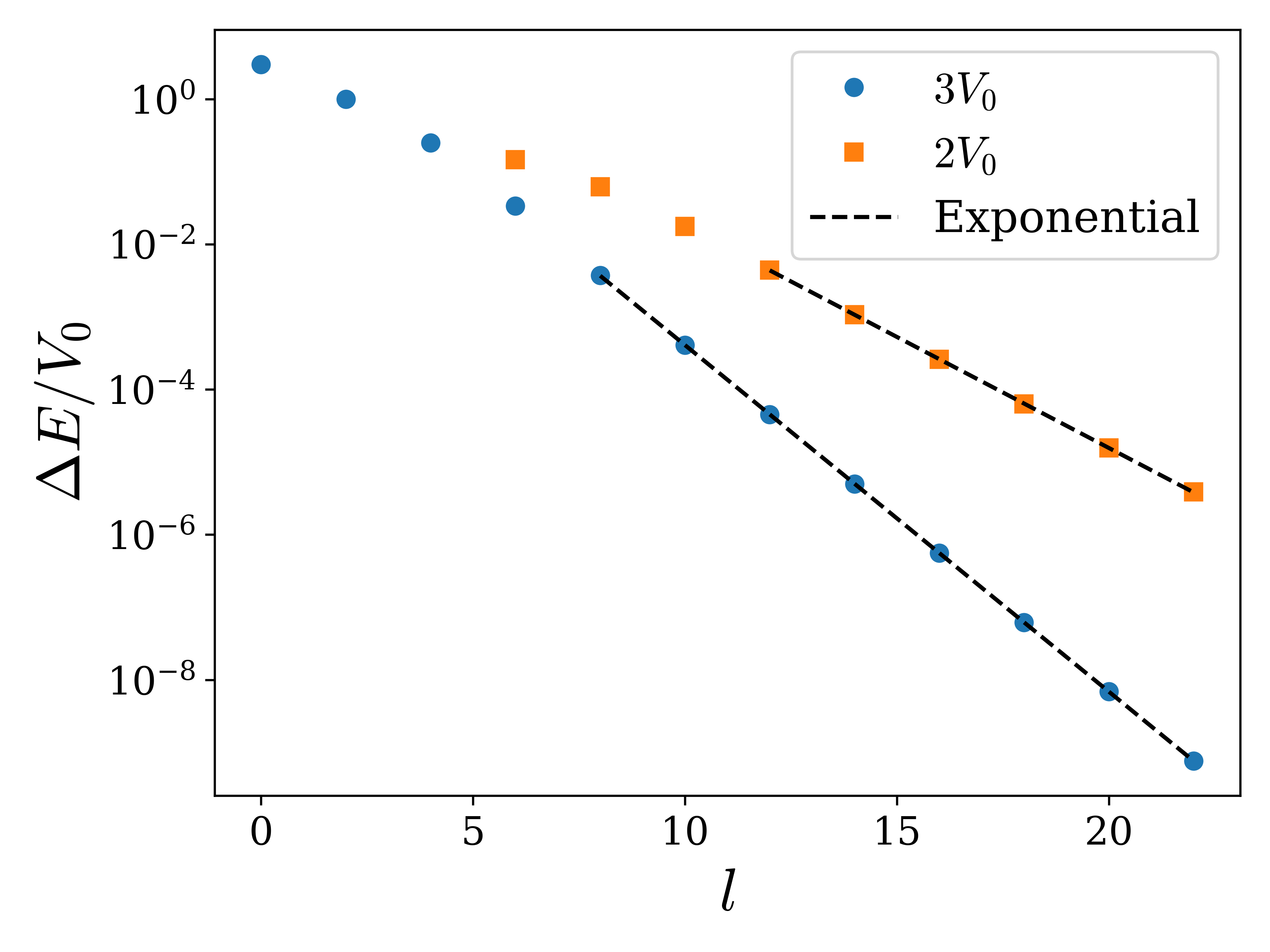}
    \caption{The series of states as functions of CoM angular momenta $l$ converging to $3V_0$ and $2V_0$ with $N=4$ in an infinite system. Their energies approach the asymptotic values exponentially (linearly on a log plot). The black lines are straight lines to guide the eye.}
    \label{fig:expo}
\end{figure}

This classical interaction energy also manifests itself in the exact quantum spectrum.
ED reveals that, for each independent cluster energy $E$ in equation~\ref{eq:decomp}, the peak in the DoS in fact comprises of a series of states with energies approaching $E$ exponentially.
As an analytic example, in an infinite boundary-less $N=3$ system, the non-zero energies appearing at angular momentum $l$ are given by $[1+2(-1/2)^l]V_0$ for $l\neq 1$.
The energy difference to $V_0$ halves for each subsequent state~\cite{mashkevich_exact_2007}.
For larger numbers of particles, the ratios between subsequent energy differences are no longer exactly the same but still stay approximately constant; see Fig.~\ref{fig:expo}.
In finite systems, there is necessarily a cutoff set by the system size, since clusters cannot be further apart than the system size itself.

To explain this exponential trend, we invoke the semiclassical approach of Bohr--Sommerfeld quantisation.
In the LLL, this means that the area of closed orbits $r^2/\ell_B^2\propto n, \text{with } n\in\mathbb{N}$.
Therefore, the quantised interaction energies are $\sim \Delta_0\ee^{-\alpha n}$ for some constant $\alpha$ and $\Delta_0$, correctly predicting the exponential series of energies.

The quantisation of the orbit area can also be related to the quantisation of angular momentum $l$ by noting that, in boundary-less systems
\begin{equation}\label{eq:cluster_size}
    \langle r^2/\ell_B^2 \rangle=2+\frac{2l}{N}.
\end{equation}
For our purposes, $l$ should be interpreted as the angular momentum in the centre-of-mass (CoM) frame.

In summary, interactions produce corrections to equation~\ref{eq:decomp}.
The eigenstates form series around $\{n_c\}$ and can be labelled with their CoM angular momentum $l$ as $\ket{\{n_c\},l}$. 
Their energies are given by $E_{\{n_c\},l}-E_{\{n_c\}}\sim \Delta_{0, \{n_c\}}\ee^{-\alpha l}$.

For later use, we compute the matrix elements $\braket{\{n_c\},l|\hat x^2|\{n_c\},l}$ and $\braket{\{n_c\},l|\hat x^2|\{n_c\},l+2}$.
The former is analytically given by equation~\ref{eq:cluster_size} and $\langle x^2\rangle=\langle r^2\rangle/2$ (with a constant offset on a torus).
The latter was numerically found to be very closely linear in $l$, too.
The dependence is linear because the matrix element should be proportional to the sizes of the clusters.

\subsubsection{The overlap of the initial state with cluster states}\label{sec:overlap}
The semiclassical theory set out above thus accounts for the states at non-integral multiples of $V_0$ in energy and predicts their properties, such as their sizes.
Returning now to the dynamical problem, in order to analytically predict the dynamics of the cloud strip, we also need to know the weights of different cluster states in the initial state.


The total weight in the cluster series described with $\{n_c\}$ can be computed by taking the overlap between the initial state and $\ket{\{n_c\}, \{\mathbf{r}_{c,i}\}}\propto \prod_c\prod_{i=1}^{n_c} \hat\psi^\dagger(\mathbf{r}_{c,i})^{c}\ket{0}$ where $\mathbf{r}_{c,i}$ are the positions of the clusters.
Also denote with $n=\sum_c n_c$ the total number of clusters.
In particular, the predicted weight of the $N$-body peak in a $N$-particle system can be shown to be exactly $(2\sqrt{\pi}/L_y)^{N-1}\sqrt{N}$ by taking $n_N=1$.
For $n>1$, the total weight in each cluster series is predicted to be
\begin{equation}\label{eq:weight}
    |w_{\{n_c\}}|^2=\frac{N!}{\prod_c n_c!\,(c!)^{n_c}} \biggl(\frac{2\sqrt{\pi}\ell_B}{L_y}\biggr)^{N-n} \sqrt{\prod_c c^{n_c}},
\end{equation}
which is accurate in the low-density limit to leading order in $1/L_y$.
This weight must be slightly modified in the thermodynamic limit where $\nu_\text{1D}$ is held constant and $L_y\to\infty$, as discussed in Appendix~\ref{sec:app_weight} along with its derivation.

Equation~\ref{eq:weight} is also intuitive: the $(1/L_y)^{N-n}$ factor accounts for the fact that, while $n$ bosons can be placed arbitrarily, each additional boson after that has only a probability $\propto 1/L_y$ to be placed at one of those positions to form clusters.
This formula predicts the formation of $\sqrt{2\pi} N\nu_\text{1D}\ell_B$ pairs in the initial state in the thermodynamic limit (correctly conserving energy), which is another intuitive result since the probability of two particles overlapping at random is $\propto \nu_\text{1D}\ell_B$.
These results apply to leading order in $\nu_\text{1D}$.

\begin{figure}[ht]
    \centering
    \includegraphics[width=0.75\linewidth]{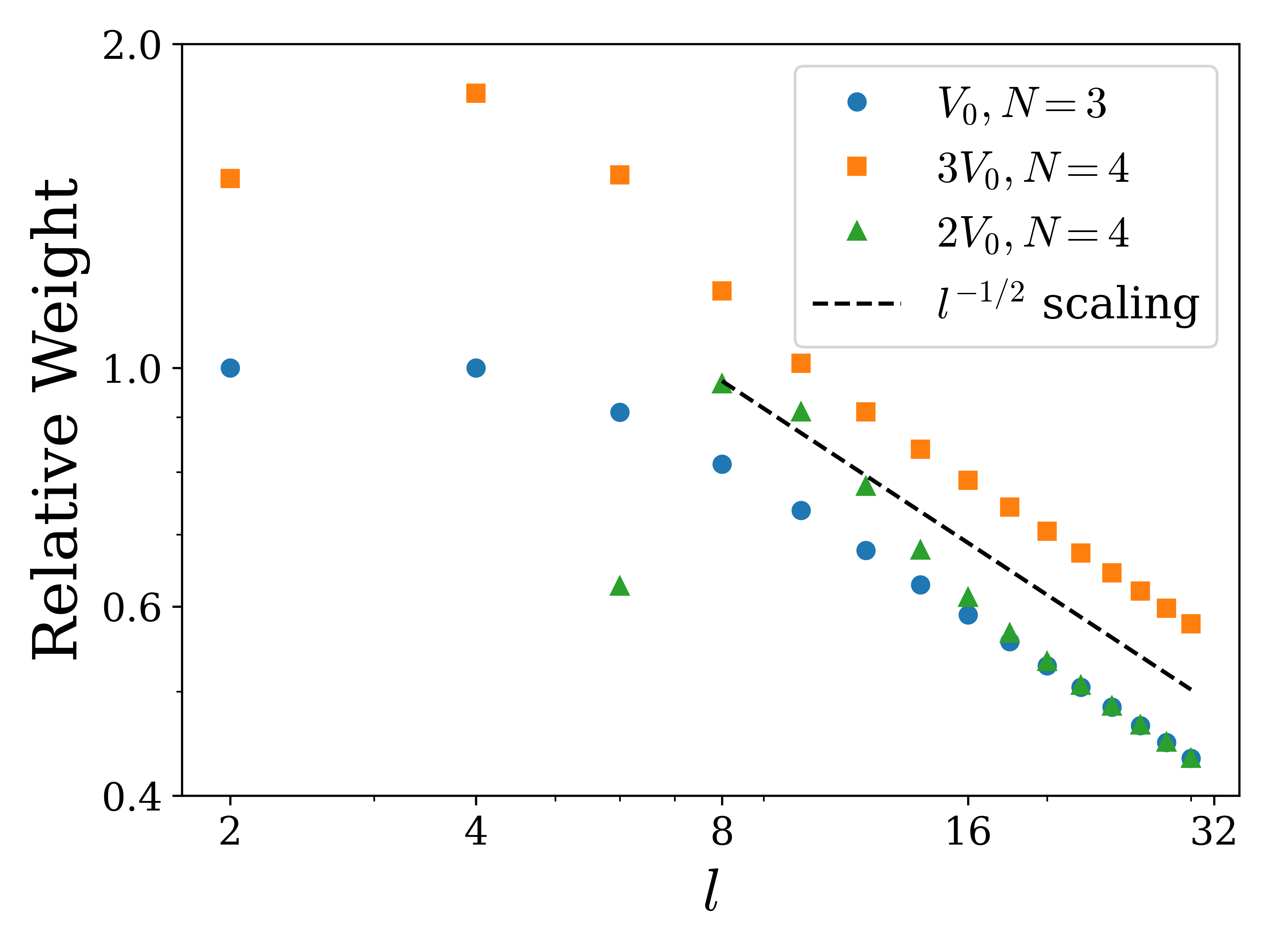}
    \caption{The relative weights of cluster states at angular momenta $l$ obtained numerically. Log-log scales are used. We see that the $V_0$ series in $N=3$ and the $3V_0$ and $2V_0$ series in $N=4$ all eventually approach a $l^{-0.5}$ scaling. The black line is to guide the eye.}
    \label{fig:weight}
\end{figure}

Within each cluster series, the weights as a function of angular momentum can be obtained numerically.
Doing so produces the $l$-dependence of relative weights in the $V_0$ series in $N=3$ and $2V_0,3V_0$ series in $N=4$ shown in Fig.~\ref{fig:weight}.
The relative weight decreases approximately as $1/\sqrt{l}$ for sufficiently large $l$.
This can be understood as follows: the initial state produces a uniform distribution of distances $r$ between clusters, but $l\sim r^2$, so the distribution of $l$ is $\sim 1/\sqrt{l}$.

When the cluster sizes become comparable to either $L_x/N$ or $L_y/N$, these predictions break down.

\subsubsection{Spreading dynamics}\label{sec:growth}
We can now explain the actual spreading dynamics of the cloud.
The initial state is given by an expansion of the form
\begin{equation}
    \ket{\Psi(t)}=\sum_{\{n_c\}}\sum_l w_{\{n_c\},l} \ee^{-\ii E_{\{n_c\},l} t}\ket{\{n_c\},l}.
\end{equation}
Calculating the time-averaged squared width then yields
\begin{multline}\label{eq:sinc}
    \frac1T\int_0^T\braket{\Psi(t)|\hat x^2|\Psi(t)}\,\mathrm{d}t = \sum_{\{n_c\}}\Biggl\{\sum_{l=0}^{L} \braket{l|\hat x^2|l}|w_l|^2 +\\
    2\sum_{l=0}^{L-2}\braket{l|\hat x^2|l+2}w_l w_{l+2}\operatorname{sinc}\biggl(\frac{1-\ee^{-2\alpha}}{\ee^{\alpha l}}\Delta_0T\biggr)\Biggr\},
\end{multline}
where the $\{n_c\}$-dependence inside the curly braces is suppressed.
Note that we have only retained terms that are diagonal in $\{n_c\}$.
Off-diagonal terms that involve matrix elements between $\{n_c\}$'s with different energies lead to fast oscillations and quickly integrate to zero.
Off-diagonal terms that involve matrix elements between $\{n_c\}$'s with the same energy are forbidden by the selection rules, as $\hat x^2$ is a one-body operator.
Extra terms involving these matrix elements will appear when the long-term behaviour of two-body operators is considered.

This sum of $\operatorname{sinc}$ functions with exponentially different frequencies produces the periodic structure seen in Fig.~\ref{fig:width_sinc}: the main features of these functions are equally spaced in logarithmic time.

The presence of a maximum $L$ in this formula is a finite size effect; $L$ depends on the size of the initial cloud as $L\sim (L_y/\ell_B)^2$.
Taken at face value, this gives the saturation in cloud width in Fig.~\ref{fig:width_sinc}.
One may object that angular momentum is not conserved on a torus, so the cloud should keep spreading.
However, the typical size of the cloud after time evolution is $\sim \sqrt{L_y\ell_B} \ll L_y$, so boundary effects are exponentially suppressed and hardly felt.

The semiclassical theory thus predicts that the cloud strip grows by gradually losing coherences between different interacting states, and it saturates at a width given by the incoherent sum over the different angular momenta.
For $N=3$, plotting the prediction with equation~\ref{eq:sinc} using calculated weights and matrix elements in the $\{n_1=1,n_2=1\}$ series gives the red curve in Fig.~\ref{fig:width_sinc}, showing excellent agreement with numerical data.

The scaling of the growth rate is given by the following analysis.
The matrix elements were numerically found to scale as $\braket{l|x^2|l+2}\sim l$, and we argued that the weights follow $|w_lw_{l+2}|\sim 1/\sqrt{l}$.
We can think of the $\operatorname{sinc}$ functions as $1$ for $\Delta_0T\ll \ee^{\alpha l}$ and $0$ for $\Delta_0T\gg \ee^{\alpha l}$, with the transition happening when $l\sim \ln T$.
Performing the sum up to $\ln T$ gives $\overline{\langle \hat x^2\rangle}(T)\sim (\ln T)^{1.5}$.
The green curves in Fig.~\ref{fig:width_sinc} show this scaling, agreeing very well with the simulation data.

\section{Long-time Dynamics and Thermalisation}
When considering the thermalisation of the gas cloud, one expects the ultimate long-time state to be a maximal entropy state for large numbers of particles.
This is because the classical dynamics of guiding centre drift is chaotic for $N\geq 4$ particles \cite{botero_two-dimensional_2016}, so one expects the long time dynamics to be ergodic.
When the volume accessible to the gas is large, the maximal entropy state at fixed energy density is a state with one ``mega-cluster'' with all of the energy and many singlets freely distributed in space.
To carry all the initial energy $E\sim \nu_\text{1D} NV_0$, this mega-cluster must have a size $N_\text{f}\sim \sqrt{\nu_\text{1D} N}$.
The remaining $N-N_\text{f}$ particles freely explore the phase space and provide the entropy.
This is analogous to the distribution of Onsager vortices, which shows clustering with diffuse tails \cite{reeves_turbulent_2022}.
Numerically, this was confirmed by noting that $\langle \hat{C}_E(0) \rangle$ at energies with multiple cluster decompositions (such as $3V_0$ in $N=6$ can be made of $n_2=3$ or $n_1=3,n_3=1$) are much more concentrated around the mega-cluster values (such as $7V_0^{(2)}$ in this example).
See Appendix~\ref{sec:app_merge} for more details.
GP simulations also qualitatively bear this out by showing the coalescence of cloud fragments.

During thermalisation, one may think of smaller clusters gradually merging into larger ones, eventually resulting in the mega-cluster with all the energy of the system.
At each step, due to the large number of particles that need to be rearranged, the time required for the merger of $c$-clusters on a length scale of $R$ can be estimated to be $\sim \ee^{cR^2}$, with some numerical factors in the exponent.
This can be understood as a Fermi Golden Rule type result from the matrix element between different cluster decompositions.
The rate-limiting step in the entire thermalisation process is the last step: the formation of an $N_\text{f}\sim \sqrt{N\nu_\text{1D}}$-cluster on a scale of $R\sim L_y$, taking $\ee^{\sqrt{N\nu_\text{1D}} L_y^2}$ time.

The logarithmic growth of observables and rates of thermalisation presented in our study is very different from the power law dependence predicted in Ref.~\cite{zerba_emergent_2025} based on a model of fracton hydrodynamics.
The reason is that the geometries are very different: our system involves a strip condensate expanding into the vacuum, whereas systems of fixed density were studied in Ref.~\cite{zerba_emergent_2025}.
To recover our results, one should allow for a density-dependent diffusion coefficient in the hydrodynamic description, and in particular one that decays exponentially as density lowers (which can be developed along the same lines as our classical cluster interaction picture).
This would then predict a ``hydrodynamics'' with our logarithmic timescales.

\section{Comparison with Gross--Pitaevskii Theory}
It is instructive to compare the ED results with the time-dependent GPE.
It is known that the GP theory is exact when the filling fraction tends to infinity while the interaction energy remains finite \cite{lieb_derivation_2006}.
For the strip-like condensate, the limits are $\nu_\text{1D}\to\infty$, and $gN/(L_y\ell_B)=g\nu_\text{1D}/\ell_B^2\to\text{constant}$.

One can immediately note that, since $g/\ell_B^2 \sim 1/\nu_\text{1D}\to 0$ in these limits and $V_0\propto g/\ell_B^2$ sets the energy scale of the dynamics of the clusters, in the GP regime, there must be no signature of the microscopic clusters of the form we found in exact diagonalisation.
This is unsurprising because the initial particle density $N/(L_y\ell_B)$ merely determines the timescale of the GP dynamics.
This was confirmed numerically.

\begin{figure}[ht]
    \centering
    \includegraphics[width=0.75\linewidth]{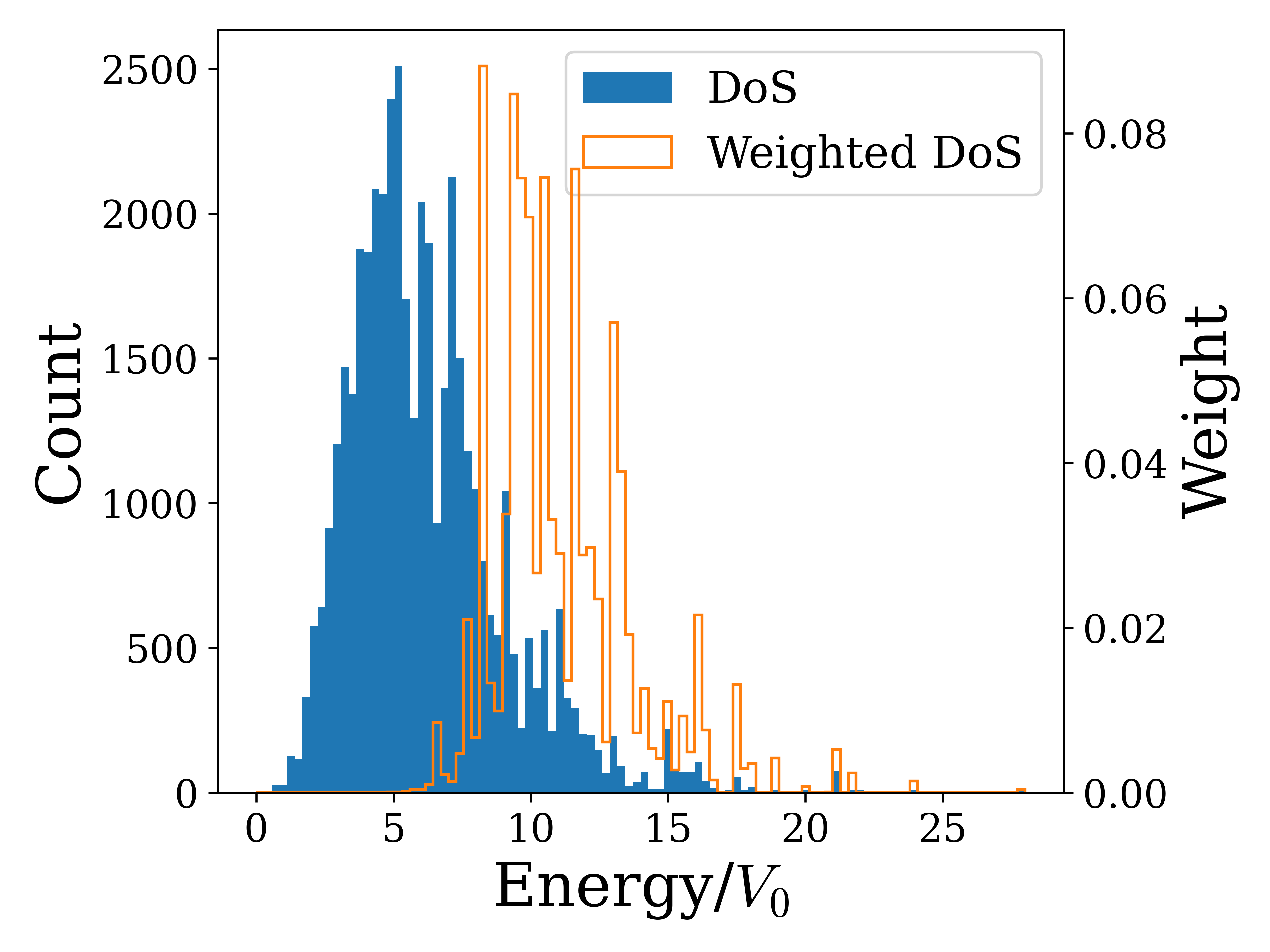}
    \caption{The ED DoS in $N=8,N_\phi=17,L_y=4\pi\ell_B$, similar to Fig.~\ref{fig:dos}. Peaks are all smeared out due to interactions. The DoS increases with energy below $\approx 5V_0$ because the number of possible positional arrangements is limited on the small torus for states with a large number of clusters like $\{n_1=8\}$.}
    \label{fig:smear}
\end{figure}

\begin{figure*}[ht]
    \centering
        \includegraphics[width=0.75\linewidth]{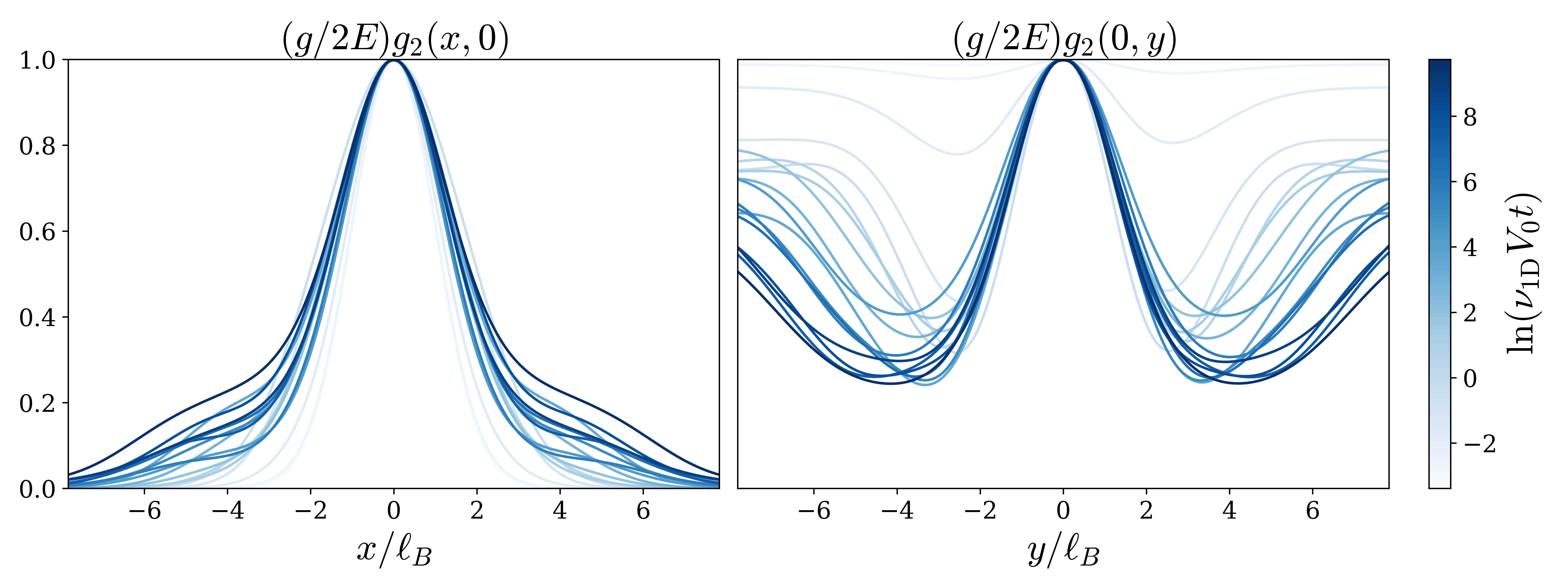}
    \caption{Sections of $g_2(\Delta\mathbf{r})$ for $N=4, N_\phi=47, L_y=6\pi\ell_B$ along the $x$ and $y$ directions as functions of time. The correlations are normalized by the total energy. Darker colors indicate later times on a logarithmic scale, as shown in the color bar. The $x$ cross-section starts as a Gaussian and expands outwards in time. The $y$ cross-section starts as a constant, develops minima as predicted by Bogoliubov theory, and plateaus at large $\Delta y$. The minima also expands outwards with time.}
    \label{fig:corr}
\end{figure*}

Consider the evolution from low density (where the cluster picture applies) to the regime of high density (where GP theory is accurate).
One expects this to be a gradual crossover as behaviours such as the fragmentation of the condensate persist into the GP regime.
We may attempt to estimate the densities at which the crossover happens.
Numerically, with an $N=8$ system, we obtained a characteristic GP frequency of around $\approx 0.063\times gN/(L_y\ell_B)$.
Since the GP frequency scales as $\nu_\text{1D}$, they become comparable to the lowest cluster frequencies of order $V_0$ when the density becomes $\nu_\text{1D}\approx 1$.

One sees a hint of how the cluster picture breaks down as the density increases in Fig.~\ref{fig:smear}, which shows the density of states for a system with $N=8, \nu_\text{1D}\approx 0.6$.
The peaks at lower energies $E\leq 5V_0$ become completely washed out, and the identifiable peaks at $6V_0$ and $7V_0$ are also broadened.
This can be qualitatively understood as the peaks being broadened by interactions and eventually overlapping.


\section{Experimental Signatures}
We are now in a position to discuss the prospects of experimentally observing the beyond-mean-field signatures of the clusters.
The most natural experimental probe employed in current experiments is the quantum gas microscope, with which one can take snapshots of the positions of all atoms with single-atom resolution.
This enables, for instance, direct measurements of the time-evolution of the squared width $\langle x^2\rangle$ from a series of destructive measurements, all with the same initial condition.

A quantum gas microscope also straightforwardly yields the correlation functions $g_2, g_3, \dots$ through the distribution of the atoms.
Here, we study the correlation function
\begin{equation}
    g_2(\Delta\mathbf{r})\equiv \int \langle \hat\psi^\dagger(\mathbf{r}) \hat\psi^\dagger(\mathbf{r}+\Delta\mathbf{r}) \hat\psi(\mathbf{r}+\Delta\mathbf{r}) \hat\psi(\mathbf{r})\rangle\, \mathrm{d}^2r.
\end{equation}

Note that while $g_2(0)$ is usually a helpful diagnostic tool for the state of the system (\textit{e.g.}, condensed vs. non-condensed states), in our system it is a constant given by the system's energy.
We therefore have to look at the position-dependence of $g_2(\Delta {\mathbf r})$ to get more information.

Some cross-sections of $g_2$ as functions of time are shown in Fig.~\ref{fig:corr}.
Several qualitative comments can be made.
\begin{itemize}
\item The gradual widening of $g_2(\Delta x,0)$  corroborates the general growth of the cloud width.
\item The initial development of minima in $g_2(0, \Delta y)$ agrees with existing experimental findings and the Bogoliubov analysis \cite{mukherjee_crystallization_2022}.
The plateaus at large $\Delta y$ can be cast in the light of independent clusters---the clusters do not care about their mutual separations once they form.
\item The outward expansion of the minima in $g_2(0, \Delta y)$ at later times is a sign of both the growth of the cluster and their separations.
Indeed, one expects that at late times the clusters become well-separated in space and are easily identifiable in quantum gas microscopy.
\end{itemize}

Additionally, while not obvious from the cross-sections in logarithmic time, the peak around $\Delta\mathbf{r}=0$ has deformations (originating from the initial anisotropy) that rotate, producing the same frequencies as described in section~\ref{sec:obs}.
Hence, $g_2(\Delta {\mathbf r})$ can also reveal information about the clusters.

Nevertheless, a key difficulty in the experimental detection of the beyond-mean-field signals is that $V_0$ is small in current systems.
The original experiment \cite{mukherjee_crystallization_2022} used a scattering length of $a_\text{s}=3.3\,\text{nm}$ and $M=3.82\times 10^{-26}\,\text{kg}$ for sodium atoms, giving an interaction strength of $g=\frac{4\pi\hbar^2 a_\text{s}}{M}\approx 10^{-50}\operatorname{J\, m^3}$.
The trap had an axial frequency $\omega_z=2.8\omega$, $\omega=2\pi\times88.6\operatorname{Hz}$, so the effective 2D interaction strength is $g_\text{2D}=3.6\times 10^{-45}\operatorname{J\,m^2}$.
The experiment also had a magnetic length of $\ell_B=1.6\operatorname{\mu m}$.
Hence, we have $V_0=\hbar\times 1.1\operatorname{Hz}$.
This means that the period of oscillations associated with the cluster energy $V_0$ is $T={2\pi\hbar}/{V_0}= 5.9\operatorname{s}$, which could be difficult to achieve.

Doubling both the interaction strength and the rotation speed would reduce it by a factor of four.
More interestingly, using the theory of clusters, we can consider higher frequency peaks, say $10V_0$ (\textit{e.g.}, between a quintuplet and 5 singlets).
According to the selection rules, this could be observed with the correlation function $g_5$. While a full measurement of $g_5$ would be very laborious, one can extract only relevant information by measuring the fifth moment of observables, such as $\langle (\hat a_k^\dagger\hat a_k)^5\rangle$.
Combining these measures gives a period of $\approx 150\operatorname{ms}$, which is more accessible in experiments.

Knowing the states are clusters, one could also envisage applications such as spectroscopically addressing precisely $n$-body clusters.
For instance, one potential way to address the pairs with RF pulses is to use scattering lengths' dependence on the hyperfine states of the atom.
Since $V_0\propto a_\text{s}$, the cluster energy of two atoms in hyperfine state $F$ is different from two atoms in state $F'$ by up to $\sim 0.6V_0$ \cite{tiesinga_spectroscopic_1996, egorov_measurement_2013}.
This means that, when using an RF pulse to drive the $F\to F'$ transition, the atoms in pair states could be off-resonance and stay in the $F$ state.
By choosing $F'$ to be an untrapped state, it is possible to create a state where only pairs remain.

\section{Discussions and Conclusions}
In conclusion, we have shown that the dynamical evolution of interacting bosons in the lowest Landau level at low filling fractions can be understood in terms of largely non-interacting clusters and that observables show oscillations at a frequency $V_0$ characteristic of two-body contact interactions.
Furthermore, long-term dynamical features can be understood with semiclassical interactions between pairs of clusters.

From our considerations of the strip-like condensate, three different processes and timescales emerge.
Firstly, when the filling fraction $\nu_\text{1D} \gtrsim 1$, the mean-field GP timescale $gn_\text{2D}$ is the fastest and can be thought of as the internal timescale of large clusters with typical particle number $\nu_\text{1D}\ell_B$.
However, when $\nu_\text{1D}\ell_B\lesssim 1$, mean field theory breaks down and this timescale becomes irrelevant.
Instead, there is the fast timescale $\tau$ given by $V_0 \tau \sim 1$ corresponding to the internal motion of (small) clusters.
Secondly, there is the timescale of interactions between pairs of clusters, which produces their motions.
Motions on the lengthscale $R$ have timescales $\ln V_0 \tau \sim R^2$ up to some constant power, giving us the spreading dynamics we explored.
Lastly, there is the timescale of the merger of clusters and of thermalisation.
Clusters of size $c$ separated by a lengthscale $R$ merge on the timescale $\ln V_0 \tau \sim cR^2$ (subject to energy conservation), so clusters merge on increasingly large scales as time goes on.
By examining the number and mean separation of $c$-clusters with quantum gas microscopy, this aspect of the thermalisation of clusters could be probed.
Full thermalisation takes a time of $\ln V_0 \tau \sim \sqrt{N\nu_\text{1D}} L_y^2$.
This, of course, corresponds to an exceedingly long time, so in realistic scenarios, the clusters may not merge fully.

Interesting connections can also be made with quantum many-body scars.
The clusters can be viewed as classical bound states that are stable over time $T\gg \hbar/V_0$. These are thus reminiscent of the ``unstable classical periodic orbits'' central to quantum scarring \cite{pizzi_genuine_2025, turner_weak_2018, serbyn_quantum_2021}.
Indeed, the behaviour we saw in logarithmic time speaks to the clusters' robustness.
The almost classical motion of the clusters also bears resemblance to the ``quantum trails'' in phase space \cite{pizzi_quantum_2025}.

Our work also invites new questions, such as how to address and manipulate the clusters in the LLL and how to understand exchange effects between the clusters.
It is also interesting to study the behaviour of the bosons on a lattice, similar to the Harper--Hofstadter model, since the required densities and interaction strengths could be easier to achieve in optical lattices \cite{leonard_realization_2023}.

\begin{acknowledgments}
We thank Zoran Hadzibabic, Claudio Castelnovo, and Andrea Pizzi for helpful discussions. This work was supported by the EPSRC [grant numbers EP/V062654/1 and EP/Y01510X/1] and a Simons Investigator Award [Grant No. 511029].
\end{acknowledgments}


\bibliographystyle{apsrev4-2}
\bibliography{cai}

\include{appendix}

\end{document}

%% file: appendix.tex
\appendix
\section{Exact two-body solutions}
While the system does not permit exact analytic solutions in general, some can be found in special cases or with suitable approximations.

We first present the exact solution of the two-body problem in the Landau gauge.
For simplicity, instead of periodic boundary conditions along the $x$-direction, we instead let the system extend to infinity by letting $N_\phi\to\infty$.
As alluded to earlier, all basis states must have the same total momentum as the initial state, namely zero.
For $N=2$, this means that all basis states are of the form $\ket{k,-k}$.
The Hamiltonian (equation~\ref{eq:ham}) has matrix elements
\begin{multline}
    \braket{k,-k|\hat H|k',-k'}=\frac{g}{2\sqrt{2\pi} L_y}\ee^{-k^2-(k')^2}\times\\
    [1+(\sqrt2 - 1)\delta_{k,0}][1+(\sqrt2 - 1)\delta_{k',0}].
\end{multline}

The two factors involving delta functions account for the boson statistics in the $\ket{0,0}$ state.
By observation, the Hamiltonian is of the form $H=\frac{g}{2\sqrt{2\pi} L_y}\mathbf{v}\mathbf{v}^\intercal$, where $\mathbf{v}^\intercal=(\sqrt{2}, 2\ee^{-(\Delta k)^2}, 2\ee^{-(2\Delta k)^2},\dots)$ in the basis $\{\ket{0,0},\ket{\Delta k,-\Delta k},\ket{2\Delta k,-2\Delta k},\dots\}$.
The factor of $2$ comes from the double counting $\ket{k,-k}=\ket{-k,k}$.
Such a matrix has rank one, and it has eigenvector $\mathbf{v}$ with eigenvalue $\frac{g}{2\sqrt{2\pi} L_y}v^2$.
All vectors orthogonal to $\mathbf{v}$ are also eigenvectors and have eigenvalue zero.
The non-zero eigenvalue can be computed by turning the sum into an integral, yielding precisely a Haldane pseudopotential $V_0=g/4\pi$.

\section{Weights in cluster series}\label{sec:app_weight}
Here we provide a derivation of equation~\ref{eq:weight}.

As described in the main text, the total weight in a given cluster configuration is given by
\begin{equation}
w=\int\frac{\braket{\Psi_0|\{n_c\},\{\mathbf{r}\}}\braket{\{n_c\},\{\mathbf{r}\}|\Psi_0}}{\braket{\Psi_0|\Psi_0}\braket{\{n_c\},\{\mathbf{r}\}|\{n_c\},\{\mathbf{r}\}}}\,\mathcal{D}\mathbf{r},
\end{equation}
where $\Psi_0\propto (a_0^\dagger)^N \ket{0}$ is the initial state, $\ket{\{n_c\}, \{\mathbf{r}\}}\propto \prod_c\prod_{i=1}^{n_c} \hat\psi^\dagger(\mathbf{r}_{c,i})^{c}\ket{0}$, and $\mathcal{D}\mathbf{r}$ indicates an integration over all positional configurations.

We see that in the dilute limit, $\braket{\{n_c\},\{\mathbf{r}\}|\{n_c\},\{\mathbf{r}\}}\sim \frac{1}{(2\pi)^N}\prod_c (c!)^{n_c}$ because the leading contribution comes from configurations where the $\mathbf{r}_{c,i}$ are all distinct.
The factors of $2\pi$ arise because the field operators projected onto the LLL are not normalized.
Similarly, $\braket{\Psi_0|\Psi_0}=N!$.
The numerator is easily evaluated using Wick's theorem, yielding
\begin{equation}
    w=\frac{(2\pi)^N N!}{\prod_c (c!)^{n_c}}\int\mathrm{d}^2r_1\,\mathrm{d}^2r_2\dots |\phi_0(\mathbf{r}_1)|^{2c_1}|\phi_0(\mathbf{r}_2)|^{2c_2}\dots
\end{equation}
The Gaussian integrals can be performed, resulting in $\frac{1}{(\sqrt{\pi}L_y)^N}\prod_c \Bigl(L_y\sqrt{\pi/c}\Bigr)^{n_c}$.
Now, the states created by $\hat\psi^\dagger$ are overcomplete, so to compensate for that, one requires that a $c$-cluster must have a weight of $1$ to be in a $c$-cluster.
This gives a correction of $c/2\pi$ per cluster, leading to
\begin{equation}
    w=\frac{N!}{\prod_c (c!)^{n_c}} \biggl(\frac{2\sqrt{\pi}\ell_B}{L_y}\biggr)^{N-n} \sqrt{\prod_c c^{n_c}}.
\end{equation}
Finally, when integrating over cluster configurations, we must note that different clusters (of the same size) are indistinguishable, giving the $1/\prod_c n_c!$ factor, proving equation~\ref{eq:weight}.

From this equation, we see that in the limit where $L_y\to\infty$, the $n_1=N$ and the $n_1=N-2,n_2=1$ terms alone reproduce the total energy of the system, while other terms are negligible.

In the thermodynamic limit, one needs to account for the fact that different clusters cannot overlap, or they would count as one cluster.
This produces a correction factor to the volume integrals when integrating over $\mathcal{D}\mathbf{r}$.
Alternatively, the normalization factor $\braket{\{n_c\},\{\mathbf{r}\}|\{n_c\},\{\mathbf{r}\}}$ is larger whenever two clusters overlap, reducing the overall overlap.
A back-of-the-envelope calculation produces a factor of the following type:
\begin{multline}
    L_y(L_y-\ell_B)\dots [L_y-(N-n-1)\ell_B] = \\
    L_y^{N-n}\exp\left[\sum_{k=0}^{n-1}\ln\biggl(1-\frac{k\ell_B}{L_y}\biggr)\right]\approx L_y^{N-n}\exp\biggl(-\frac{n^2\ell_B}{2L_y}\biggr).
\end{multline}
There could be numerical factors within the exponent.
One then sees that this suppresses the divergent ``probabilities'' in the limit the number of clusters $n\to\infty$ while $n/L_y$ remains constant.
Such a term does not change the predicted number of pairs in the low $\nu$ limit.

\section{Merger of clusters}\label{sec:app_merge}
\begin{figure}
    \centering
    \includegraphics[width=0.75\linewidth]{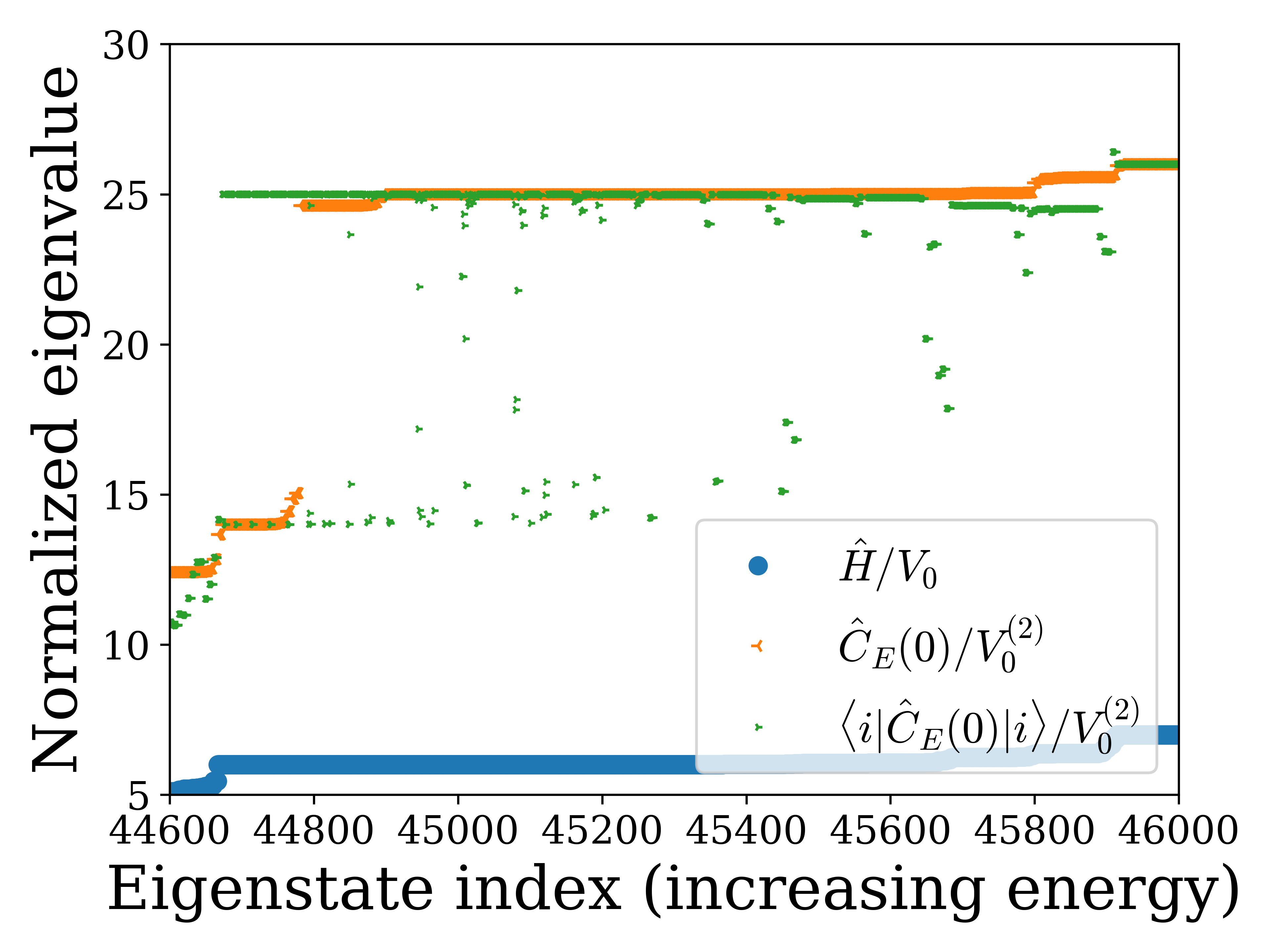}
    \caption{Eigenstates in a six-particle system around $E=6V_0$. Plotted quantities are the same as in Fig.\ref{sfig:eigval}. At this energy, the state could be comprised of two triplets or one quadruplet and two singlets. The former leads to $\langle \hat{C}_E(0)\rangle = 14V_0^{(2)}$, the latter $\langle \hat{C}_E(0)\rangle = 25V_0^{(2)}$. Both are seen, and the quadruplet state is much more common, as expected. More importantly, the presence of states with $\langle \hat{C}_E(0)\rangle$ in between those two values indicate mixing of the cluster decompositions in the energy eigenstates. Correspondingly, there is a small shift in $\langle \hat{C}_E(0)\rangle$ for the higher energy states.}
    \label{fig:mix}
\end{figure}

One way to study the merger of clusters is through the operator $\hat{C}_E(0)$ as we did in Fig.~\ref{fig:dos}.
An example for $N=6$ is shown in Fig.~\ref{fig:mix}.

This $N=6$ system differs from the $N=4$ system previously considered in that there are multiple possible cluster decompositions $\{n_c\}$ at the same energy.
Consequently, instead of the one-to-one correspondence between the energy and $\langle C_E(0)\rangle$, we see a mixture between the different possibilities.
This is an indication that, subject to energy conservation, clusters will merge and change their identities eventually.

The higher energy states (higher index) shown in the figure consist of clusters that are closer together, and as described in the main text the matrix element between different decompositions are greater.
The mixing between decompositions appears quite thorough for the high-energy states.
As the clusters become increasingly separated (higher CoM angular momentum), the matrix element falls off as a power of a Gaussian.
This means that the numerical algorithms cannot be trusted to produce the true eigenstates, and numerically we obtain states at the unmixed $\langle \hat{C}_E(0)\rangle$ values.
We still expect the clusters to merge even for these eigenstates despite the smallness of the matrix elements, and the eigenstates will obey the eigenstate thermalization hypothesis based on the very general arguments laid out in the main text.